\documentclass[aps,prB,twocolumn,reprint,superscriptaddress]{revtex4}

\usepackage{hyperref}
\usepackage{graphicx}  
\usepackage{bm}        
\usepackage{amssymb}   
\usepackage{xspace}
\usepackage{verbatim}
\usepackage{color}
\usepackage{soul}
\usepackage{subfigure}
\usepackage[export]{adjustbox}
\usepackage{dcolumn}
\usepackage{amsthm,stmaryrd,mathtools} 
\usepackage{txfonts}
\usepackage{braket}
\usepackage{amsmath}
\usepackage{xcolor}
\usepackage{array}
\usepackage{multirow}
\usepackage{tabularx}
\usepackage{booktabs}
\usepackage{lipsum}

\begin{document}

\title{Mobility edge in long-range interacting many-body localized systems}

\author{Rozhin Yousefjani}%
\email{RozhinYousefjani@uestc.edu.cn}
\affiliation{Institute of Fundamental and Frontier Sciences, University of Electronic Science and Technology of China, Chengdu 610051, China}

\author{Abolfazl Bayat}
\email{abolfazl.bayat@uestc.edu.cn}
\affiliation{Institute of Fundamental and Frontier Sciences, University of Electronic Science and Technology of China, Chengdu 610051, China}

\begin{abstract}

As disorder strength increases in quantum many-body systems a new phase of matter, the so-called many-body localization, emerges across the whole spectrum. This transition is energy dependent, a phenomenon known as mobility edge, such that the mid-spectrum eigenstates tend to localize at larger values of disorder in comparison to eigenstates near the edges of the spectrum. 
Many-body localization becomes more sophisticated in long-range interacting systems.
Here, by focusing on several quantities, we draw the phase diagram as a function of disorder strength and energy spectrum, for a various range of interactions.
Regardless of the underlying transition type, either  second-order or Kosterlitz-Thouless, 
our analysis consistently determines the mobility edge, i.e. the phase boundary across the spectrum. 
We show that, long-range interaction enhances the localization effect and shifts the phase boundary towards smaller values of disorder. 
In addition, we establish a hierarchy among the studied quantities concerning their corresponding transition boundary and critical exponents. Interestingly, we show that  deliberately discarding some information of the system can mitigate finite-size effects and provide results in line with the analytical predictions at the thermodynamic limit. 
\end{abstract}

\maketitle
\section{Introduction}

Ergodicity principle, as the foundation of statistical physics, is violated in disordered systems. 
Many-Body Localization (MBL) is the primary example of such phenomenon in interacting disordered systems which has attracted lots of attention in recent years~\cite{Rev1,Rev2,Rev3,Rev4,Rev5,Rev6,Rev7}. 
Several features of the MBL have been characterized through static and dynamical analyses, including Poisson-like level statistics~\cite{Huse1,Huse2,Huse3,Huse4,Laflorencie1}, area-law entangled eigenstates~\cite{Area-law1,Area-law2}, logarithmic entanglement growth~\cite{Laflorencie2,UnboundedEG,EGSR1,EGSR2,EGSR3,EGSR4,EGSR5}, suppression of transport~\cite{ST1,ST2,ST3,ST4,ST5}, power-law decay of local correlations~\cite{OTOC1,OTOC2,OTOC3,OTOC4,OTOC5,OTOC6},  emergence of memory~\cite{IM1,Memory1,IM3,IM4,IM5,IM6,IM7,
Memory2}, 
and connection to topological phases~\cite{Topology1,Topology2}. 
Unlike quantum phase transition, which is a property of the ground state~\cite{QPT1}, the transition from ergodic to MBL takes place across the whole spectrum, as disorder strength increases. This makes the detection and characterization of the MBL transition very challenging~\cite{CriticalMBL,superconducting9,CriticalRbatom,
superconducting8,Jakub1,challeng1,challeng2,
challeng3,challeng4}.
In fact, each energy eigenstate localizes at a different disorder strength, a phenomenon called mobility edge which has been explored theoretically~\cite{Laflorencie1,Area-law1,Mobility1,Mobility2,Mobility3,Mobility4,
Mobility5,Mobility6,Mobility7,Mobility8,Mobility9,
Mobility10} and  observed experimentally~\cite{superconducting1}. 
Since MBL is a property of the whole spectrum, its numerical investigation is mostly limited to exact diagonalization of small systems. This makes it very challenging to extract information about the thermodynamic limit. Thanks to the recent development of quantum simulators, MBL experiments have been implemented in various platforms such as superconducting devices~\cite{superconducting1,superconducting2,
superconducting3,superconducting4,
superconducting5,superconducting6,
superconducting7,superconducting8,
superconducting9}, optical lattices~\cite{coldatoms1,coldatoms2,coldatoms3,coldatoms4,
2DRDAtoms,CriticalRbatom,Rbatom}, ion traps~\cite{Iontrap1,Iontrap2,Iontrap3},
 nitrogen-vacancy centers in diamond~\cite{nitrogenvacancy1,nitrogenvacancy2}, 
 and photonic systems~\cite{photonic}.   \\

Perhaps the most challenging problem in the MBL context is the characterization of the MBL transition and its corresponding mobility edge~\cite{Rev5,Panda2019,Morningstar2022,
Polynomially2020,Shiftinvert2018}. 
In a crucial analytical contribution by Harris~\cite{HarrisBound1}, which has been extended by others~\cite{HarrisBound2,HarrisBound3},  the MBL transition has been described as a continuous second-order phase transition which is accompanied by the emergence of a diverging length scale $\xi_{_{SO}}{\sim} |W-\omega|^{-\nu}$ in the system.  
Here, $W$ is the strength of disorder, $\omega$ is critical disorder strength beyond which the system is localized, and $\nu$ is a critical exponent.
Harris analysis predicts that $\nu$ has to satisfy $\nu{\ge}2/d$ in a $d$-dimensional system.  
However, most of the conventional quantities that have been studied in 1-dimensional MBL systems, such as entanglement entropy and level statistics,  violate Harris criteria and result in $\nu {\sim} 1$~\cite{Laflorencie1,Memory1}. 
There are some exceptions, such as Schmidt gap~\cite{IM1,Schmidtgap} and diagonal entropy~\cite{IM1,DEMBL1,DEMBL2}, which either satisfy the Harris criteria or at least violate it less by giving $\nu{\sim}2$. 
There are two distinct explanations for the inconsistency of numerical simulations with the Harris bound.
Either the accessible system sizes are too small to emulate the thermodynamic limit~\cite{Panda2019,Polynomially2020},
or describing the MBL transition as a second-order phase transition is not valid and one has to explain it as a  Kosterlitz-Thouless type~\cite{KT1,KT2,KT3,KT4,KT5,
KT6,KT7,KT8,KT9}. 
The suggestion of describing MBL transition as a Kosterlitz-Thouless transition type with a diverging length scale  
$\xi_{_{KT}}{\sim} \exp{(b/\sqrt{\vert W-\omega\vert})}$, received support from real-space renormalization group approaches based on the avalanche scenario~\cite{KT1,KT2,KT3} and has been investigated numerically~\cite{KT4,KT5,KT6,KT7,KT8,KT9}. 
Nonetheless, due to small system sizes, the debate about the transition type is far from being settled.
\\

Investigation of the MBL beyond the conventional nearest-neighbor interactions may open new questions.
Long-range interactions, such as Coulomb, dipole-dipole and van der Waals, 
are of utmost importance as they naturally arise in many systems. 
On the experimental side, some of the quantum simulators, such as  ion traps~\cite{Iontrap1,Iontrap2,Iontrap3} and Rydberg atoms~\cite{2DRDAtoms,CriticalRbatom,Rbatom}, are naturally governed by long-range interactions.
The existence of an MBL phase and its principal properties  in long-range interacting systems is still under dispute~\cite{IM7,PD13,PD1,PD2,PD3,PD4,PD5,PD6,PD7,
PD8,PD9,PD10,PD11,UAEG,PD12,PD14,
Mobility8,Mobility10,Mobility6}. 
For instance, while earlier studies of the MBL in long-range systems suggested an algebraic growth of entanglement entropy~\cite{UAEG,LREG1,LREG2,LREG3,LREG4,LREG5,LREG6,LREG7}, recent work shows non-algebraic behavior through long-time numerical simulations~\cite{Yousefjani}.
Interestingly, different types of long-range couplings can have different effects on the localization issue.  
While long-range tunneling can increase delocalization, long-range (Ising) interaction enhances localization~\cite{PD1,LREG5,UAEG,Yousefjani}.  
Several important issues are still open in long-range MBL systems, including (i) the main nature of the transition   
(ii) the phase boundary, and thus the mobility edge, along the energy spectrum as the strength of long-range interaction varies;  and (iii) the energy dependence of the critical exponents, associated with the transition types, along the mobility edge. \\

In this paper, we aim to address these issues by considering several quantities, including level statistics ratio, entanglement entropy, diagonal entropy, and Schmidt gap.
For both of the transition types, i.e. second-order and Kosterlitz-Thouless transition,  we draw the phase diagram extracted from the four aforementioned quantities across the whole spectrum as the strength of long-range interaction varies.
Several results have been observed.
We show that, regardless of the transition type, all the studied quantities determine the phase boundary and the D-shape mobility edge consistently. 
This feature, has already been observed, both analytically~\cite{Laflorencie1,Area-law1,
Mobility1,Mobility2,Mobility3,Mobility4,
Mobility5,Mobility6,Mobility7,Mobility8,Mobility9}
 and experimentally~\cite{superconducting1}, for MBL transition as a continuous second-order transformation.
However, the emergence of this characteristic for the Kosterlitz-Thouless type has not been reported previously.   
Besides, we show that long-range interaction enhances the localization and shifts the mobility edge towards smaller values of $\omega$, consistent with previous studies~\cite{PD1,UAEG,PD2}. 
In addition, while in most of the literature both of the critical exponents  $\nu$ (in second-order transition) and $b$ (in Kosterlitz-Thouless transition)  
are only computed for the mid-spectrum,
we determine them along the whole spectrum.
As the main result, our analysis establishes a hierarchy
 among our quantities with respect to the phase boundary and critical exponents.
Interestingly, we show that for very wide range of long-range interactions,  Schmidt gap can give results fully consistent with the Harris criteria.
This can be explain based on this fact that 
Schmidt gap converges to its thermodynamic limit faster than the other three quantities.

\section{Model.}
We consider an open-boundary chain of $N$ spin-1/2 particles in the presence of a random magnetic field. While spin tunneling is restricted between nearest-neighbor sites, interaction between particles is taken to be long-range which algebraically decays by exponent $\alpha{>}0$. The Hamiltonian reads
\begin{equation}\label{Eq. Hamiltonian}
H = -\sum_{i=1}^{N-1}(S_i^{x}S_{i+1}^{x}+S_i^{y}S_{i+1}^{y}) - \sum_{i \neq j=1}^{N}\dfrac{1}{|i - j|^{\alpha}}S_i^{z}S_j^{z} + \sum_{i=1}^{N}h_i S_i^{z},
\end{equation}
here $S_i^{(x,y,z)}$ are the spin-$1/2$ operators for
qubit $i$, and $h_i$ denotes a random magnetic field in the $\hat{z}$ direction acting on qubit $i$ which is drawn from a uniform distribution $[-W,W]$, with $W$ being the strength of disorder. Note that, by varying $\alpha$  we can smoothly interpolate between a fully connected graph (i.e. $\alpha{=}0$) and standard  nearest-neighbor 1-dimensional chain (i.e. $\alpha{\rightarrow}\infty$).
Many types of long-range models such as Coulomb and dipole-dipole interactions are special examples of Hamiltonian $H$.
Long-range interacting systems, with tunable $\alpha$, can be realized in cold atoms in optical lattices~\cite{MagneticAtom}, ion traps~\cite{Iontrap1,Iontrap2,Iontrap3} and polar molecules~\cite{PD3}.
The isotropic Hamiltonian Eq.~(\ref{Eq. Hamiltonian}) corresponds to multipole expansion introduced in~\cite{PD3} which has a critical dimension $d_c{=}(\alpha+2)/2$ and gets localized for all $\alpha{>}0$.
The possibility of observing such consistent MBL in our model was also corroborated by other works using resonant pairs analysis~\cite{LREG5,Burin} and mean-field theory~\cite{PD1}.
Indeed, the existence of MBL in one-dimensional systems is shown through analytical analysis~\cite{Imbrie2016,ImbriePRL} as well as numerical investigations~\cite{Polynomially2020}.
Nonetheless, recent studies~\cite{Rev5,Morningstar2022,Avalanche1,Avalanche2,
Avalanche3,Avalanche4,Avalanche5,Avalanche6} raised the possibility that for some disorder ansatz formation of rare ergodic spots (randomly formed chaotic domains of reduced disorder), 
may drive an avalanche, by thermalizing nearby spins, and eventually thermalize the whole system.
The possibility of finding such rare ergodic spots increases as the system gets larger~\cite{Rev5,Avalanche1}. 
The formation of such rare ergodic spots is sensitive to the range of interaction and 
is mostly endemic to models with uncorrelated disorder and systems which support the existence of mobility edge~\cite{Mobility8}.
So, our model is no exception and avalanche effect might be observable for certain random ansatz, though we hardly encounter such problem in our numerical analysis.
This might be due to finite sizes which are available to numerical calculations as MBL is commonly observed in 1-dimensional systems for sufficiently large disorder $W$~\cite{Rev5,Avalanche1}. 
Indeed, the numerical limitations leave the fate of MBL 
with respect to the possibility of avalanche still an open problem.\\

\begin{figure}[t!]
\includegraphics[width=\linewidth]{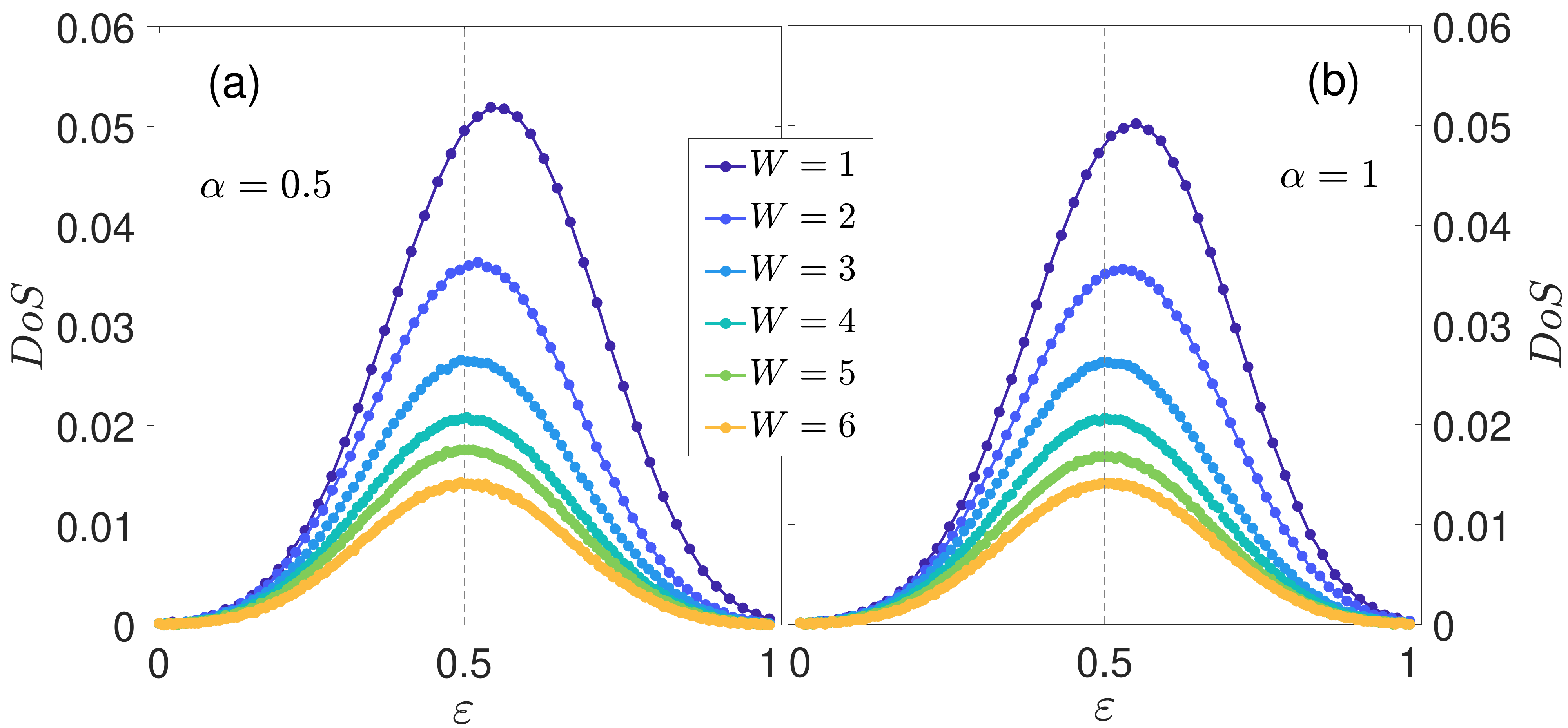}
\includegraphics[width=\linewidth]{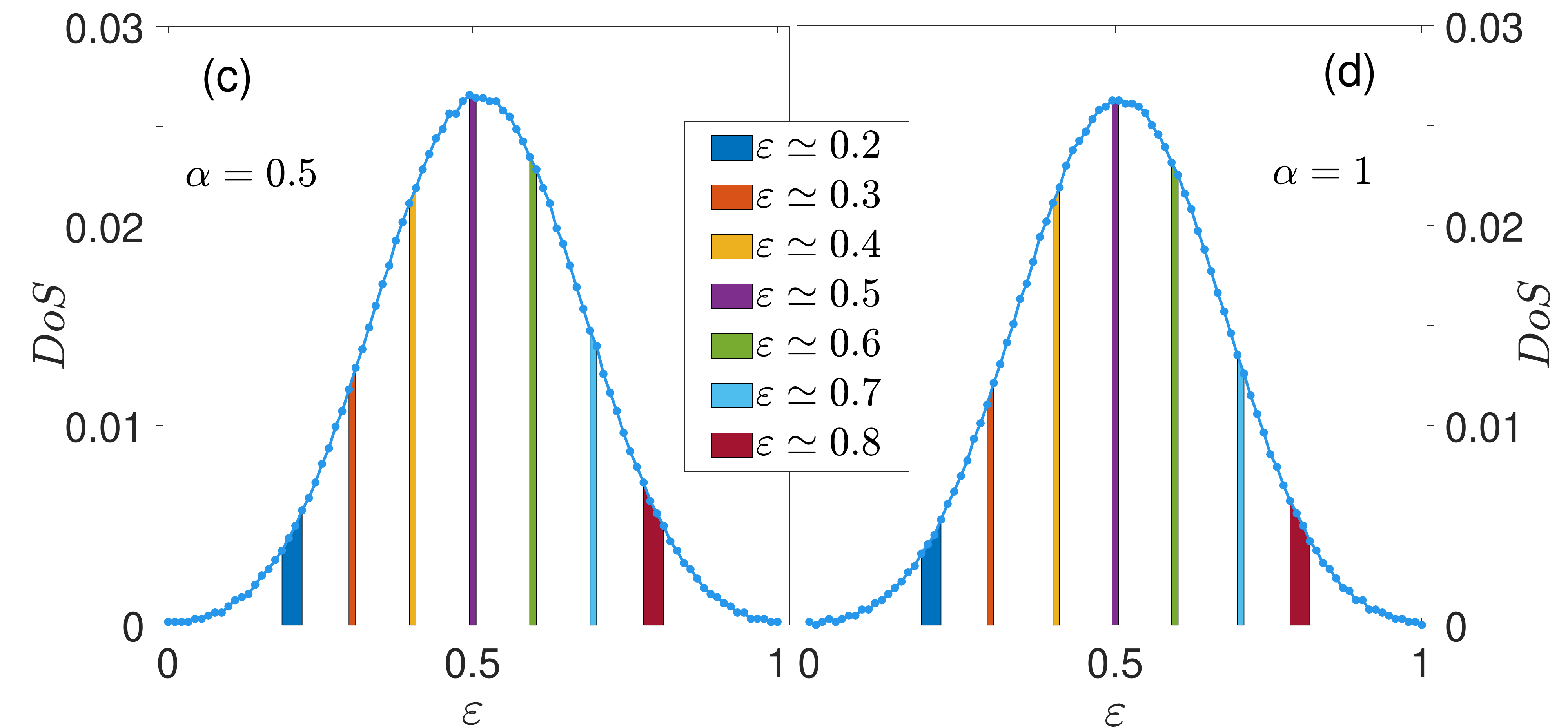}
\caption{\textbf{Upper panels}: Density of states (DoS) as a function of rescaled energy $\varepsilon$ for different disorder strengths in system with exponents (a): $\alpha{=}0.5$ and (b): $\alpha{=}1$. \textbf{Lower panels}: DoS as a function of $\varepsilon$ in system with fixed disorder strength $W{=}3$ and exponents (c): $\alpha{=}0.5$, and  (d): $\alpha{=}1$. The colored bars represent the domains from them $M{=}50$ consecutive eigenstates in the vicinity of rescaled energies  $\varepsilon{\in}\{0.2,\cdots,0.8\}$ have been selected. All the results have been obtained for a system of size $N{=}15$. }\label{fig:Dos}
\end{figure}

For our numerical analysis, we restrict ourselves to the subspaces of $S^z_{tot}{=}0$  or $S^z_{tot}{=}1/2$ (with $S^z_{tot} = \sum_{j}S^z_j$) for even and odd $N$'s, respectively. 
For systems of length $N{=}10,11,\cdots,15$, 
to achieve a good statistic, we provide $2000$ (for $N{\leqslant} 13$) to $1000$ (for $N{=}14,15$) sample realizations of the random field. 
For each set of random fields and with the means of exact diagonalization, we generate $M$ eigenstates $\{\vert E_{k}\rangle\}$ of the Hamiltonian $H$ around the rescaled energy $\varepsilon{\in}\{0.2,\cdots,0.8\}$, where $\varepsilon$'s have been calculated as 
$\varepsilon{=}(E - E_{min})/(E_{max}-E_{min})$ in which $E_{max}$ and $E_{min}$ are the extremal eigenvalues of $H$.
In Fig.~\ref{fig:Dos}, we depict the density of states (DoS) as a function of energy $\varepsilon$ in a system of size $N{=}15$ and exponents (a): $\alpha{=}0.5$ and (b): $\alpha{=}1$.
Except for small values of disorder, the DoS is symmetric around the mid-spectrum energy and its maximum hardly changes from $\varepsilon{=}0.5$. This is in sharp contrast with long-range systems with random exchange couplings in which the maximum of DoS is highly skewed towards low energy part of the spectrum~\cite{PD10,Mobility10}.        
Note that most of MBL analyzes are focused on the energy corresponding to the peak of the DoS where the spectrum is very dense. In our case, this is mostly around $\varepsilon{=}0.5$, namely mid-spectrum. For the sake of completeness, in Figs.~\ref{fig:Dos}(c) and (d), we plot DoS as a function of energy for the disorder strength  $W{=}3$ and two different values of $\alpha$, namely $\alpha{=}0.5$ and $\alpha{=}1$ respectively. 
The colored bars represent the domain from which we select $M{=}50$ consecutive eigenstates $\{E_k\}$ in the vicinity of the rescaled energies $\varepsilon{\in}\{0.2,\cdots,0.8\}$.      
For every eigenstate $\vert E_{k}\rangle$ one can compute the desired quantity which results in $\mathcal{O}_k$. 
To have an average behavior of eigenstates at energy $\varepsilon$, we average over the $M{=}50$ eigenstates around that energy which results in $\langle \mathcal{O} \rangle=1/M\sum_{k=1}^M \mathcal{O}_k$. 
In addition, to observe the effect of disordered field, one has to also average over different random samples (typically 1000-2000 samples) which is denoted by $\overline{\langle \mathcal{O} \rangle}$.

\section{Quantities}
Our first quantity is the gap ratio $r$ which is a well-established tool to study the statistical properties of the energy levels 
$\{E_k\}$ of the Hamiltonian $H$.
For energy gaps $\delta_{k}{=}E_{k+1}{-}E_{k}$, the ratio $r_{k}{=}\min(\delta_{k+1},\delta_{k}){/}\max(\delta_{k+1},\delta_{k})$ and its average $\overline{\langle r\rangle}$, over energy levels and sample realizations, can diagnose a global change in the spectral statistics. 
While in the MBL phase the energy levels are spaced according to Poisson statistics, yielding $\overline{\langle r\rangle} {\simeq} 0.3863$, in the ergodic phase the strong
repulsion between neighboring levels causes statistics that match with those of the Gaussian orthogonal ensembles, resulting in $\overline{\langle r\rangle}{\simeq}0.5307$.\\

The second quantity of interest is the entanglement entropy $S_{_{EE}}$, which is a widely used tool for quantifying entanglement between two complementary parts of a pure many-body system. 
For any given eigenstate $\vert E_k\rangle$ one can obtain the reduced density matrix $\rho^{(k)}_{_{L}}$ by tracing out $\lceil L/2\rceil$ qubits on the right side of the chain. 
Therefore, the entanglement entropy between left and right cuts of the chain is defined as 
$S_{_{EE}}(\rho^{(k)}_{_{L}}){=}{-}\mathrm{Tr}[\rho^{(k)}_{_{L}}\ln(\rho^{(k)}_{_{L}})]$. 
In the ergodic phase, the eigenstates are expected to behave like a random pure state and exhibit a volume law entanglement entropy determined by Page entropy 
$S^{P}_{_{EE}}{=}(1-\mathcal{D}_{_{L}})/2\mathcal{D}_{_{R}} + \sum_{k=\mathcal{D}_{_{R}} +1}^{\mathcal{D}_{_{L}}\mathcal{D}_{_{R}}}1/k$ with $\mathcal{D}_{_{L}}$ and $\mathcal{D}_{_{R}}$ as the Hilbert space dimensions of the left and right chain cuts~\cite{PageVNE}.
In contrast, in the localized phase due to the emergent integrability the eigenstates exhibit an area-law entanglement. 
Therefore, in deep ergodic and MBL phases, one expects to have $\overline{\langle S_{_{EE}}\rangle} /S^{P}_{_{EE}}\sim 1$ and $\overline{\langle S_{_{EE}}\rangle}/S^{P}_{_{EE}}\sim 0$, respectively, for the averaged entanglement entropy. \\

As the third quantity, we consider the diagonal entropy of a subsystem. 
Replacing $\rho^{(k)}_{_{L}}$ in $S_{_{EE}}(\rho^{(k)}_{_{L}})$ by the decohered density matrix $\varrho^{(k)}_{_{L}}$, in which all the off-diagonal elements in computational basis are set to zero, results in the diagonal entropy $S_{_{DE}}(\varrho^{(k)}_{_{L}}){=}{-}\mathrm{Tr}[\varrho^{(k)}_{_{L}}\ln(\varrho^{(k)}_{_{L}})]$. 
This quantity has recently been used in the context of MBL~\cite{IM1,DEMBL1,DEMBL2}. 
In the ergodic phase, the diagonal entropy provides an upper bound for entanglement entropy, namely $S_{_{DE}}(\varrho^{(k)}_{_{L}}) \geqslant S_{_{EE}}(\rho^{(k)}_{_{L}})$. 
However, in the deep localized phase they remain close to each other, namely $S_{_{DE}}(\varrho^{(k)}_{_{L}}) {\sim} S_{_{EE}}(\rho^{(k)})$,  as $\rho^{(k)}_{_{L}}$ and its decohered state $\varrho^{(k)}_{_{L}}$ have high fidelity. 
Analogs to the entanglement entropy, by calculating the maximum diagonal entropy for a typical random pure state as $S^{P}_{_{DE}}\simeq \ln(0.48\mathcal{D}_{_{L}}){+}\ln(2)$~\cite{PageDE1,PageDE2}, one can determine two limiting behaviors $\overline{\langle S_{_{DE}}\rangle} /S^{P}_{_{DE}}\sim 1$ and  $\overline{\langle S_{_{DE}}\rangle} /S^{P}_{_{DE}}\sim 0$ for the ergodic and localized phases, respectively. \\

The fourth and final indicator is the Schmidt gap of the reduced density matrix $\rho^{(k)}_{_{L}}$ which can be calculated by using its two largest eigenvalues, denoted by $\lambda^{(k)}_{1}$ and $\lambda^{(k)}_{2}$, as $\Delta(\rho^{(k)}_{_{L}}){=}\lambda^{(k)}_{1}{-}\lambda^{(k)}_{2}$. 
In contrast to the entanglement entropy which considers all the spectrum of $\rho^{(k)}_{_{L}}$, the Schmidt gap only focuses on  two dominant eigenvalues.
In the ergodic phase  
$\rho^{(k)}_{_{L}}$ is expected to be a thermal state with an infinite temperature (at least for mid-spectrum eigenstates) which results in 
$\overline{\langle\Delta\rangle}{\sim}0$.
In the localized phase the reduced density matrix $\rho^{(k)}_{_{L}}$ tends to become a pure state and thus the Schmidt gap increases such that in the deep localized phase, one has $\overline{\langle\Delta\rangle} \rightarrow 1$.

\section{Transition type}
One of the fundamental features of quantum phase transitions is the emergence of scale invariance in the vicinity of the transition point~\cite{QPT1}. In the context of MBL, the scale invariance implies that a length scale emerges in the system and thus  all quantities are expected to scale as~\cite{Bayat}  
\begin{equation}\label{eq:FSS-ansatz}
   \overline{\langle \mathcal{O} \rangle} = f(N/\xi), 
\end{equation}
where $N$ is the system size, $\xi$ is the emerging length scale that diverges at the transition point, and 
$f(\cdot)$ is an arbitrary function which depends on the quantity of interest.
Most of the literature characterize the MBL transition as a continuous second-order transformation which is described by a diverging length scale~\cite{Huse2,Jakub2,Bayat,Huse3,Huse4,Laflorencie1,Area-law1,Mobility1,Mobility3,Mobility4,CriticalMBL}
\begin{equation}\label{Eq.SO}
\xi_{_{SO}}{\sim}\dfrac{1}{\vert W-\omega\vert^{\nu}},
\end{equation}
where, $\omega$ is the critical disorder strength beyond which the system is localized, and $\nu$ is a universal critical exponent. 
It has analytically been shown that $\nu{\geqslant}2/d$ (known as Harris criterion), with $d$ being the dimension of the system~\cite{HarrisBound1,HarrisBound2,HarrisBound3}.
While describing  MBL transition as a second-order phase transition is supported by the phenomenological approaches  based on real-space renormalization group~\cite{CriticalMBL}, most of the numerical analyses for one-dimensional systems, with the exception of Refs.~\cite{Memory1,Schmidtgap}, give $\nu{\sim}1$ which is in contradiction with Harris bound. 
The failure of capturing the right value of $\nu$, can be associated to very finite system sizes that are accessible for numerical simulations~\cite{Memory1,Schmidtgap}.\\

Recently, a new approach to describing the MBL transition has also been adopted. 
In this approach, people argue that the very basic assumption of the second-order phase transition might be wrong, and thus the Harris criterion is totally irrelevant.
They suggest that the transition might be of Kosterlitz-Thouless type~\cite{KT1,KT2,KT3}.  
In such transition, while the scale invariance of Eq.~(\ref{eq:FSS-ansatz}) is still  valid, the emerging length scale is described as   
\begin{equation}\label{Eq.KT}
\xi_{_{KT}}{\sim}\exp{(\dfrac{b}{\sqrt{\vert W - \omega \vert}})},
\end{equation}
where $\omega$ is again the critical disorder strength beyond which the system is localized and $b$ is a non-universal fitting parameter.
Despite several attempts~\cite{KT4,KT5,KT6,KT7,KT8,KT9} for settling the type of transition, by comparing the second-order versus Kosterlitz-Thouless transitions, the finite system sizes do not allow to reach a conclusive answer. Therefore, in this paper, for the sake of completeness,  we  perform finite-size scaling analysis for both of the transition types.

\section{Finite-size scaling analysis}

\begin{figure}[t!]
\includegraphics[width=\linewidth]{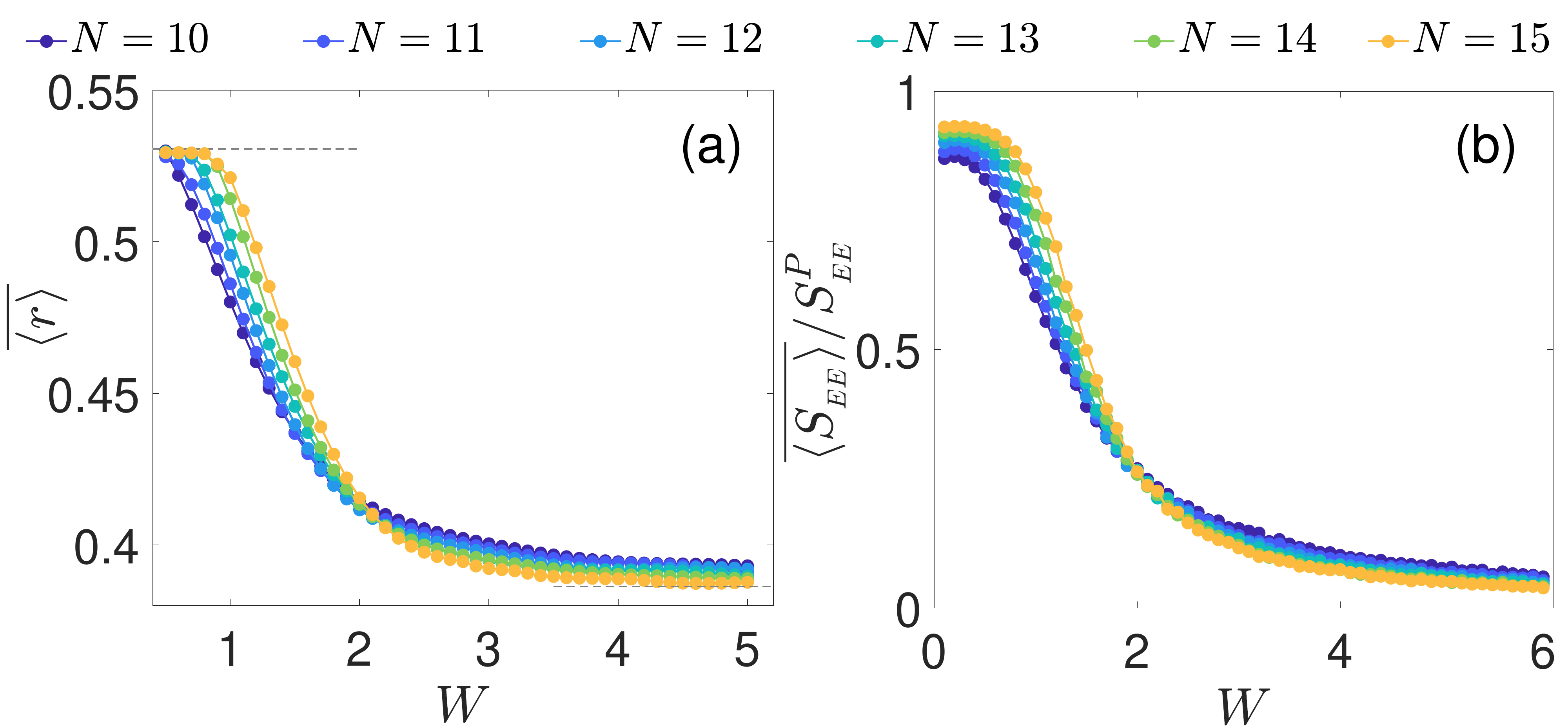}
\includegraphics[width=\linewidth]{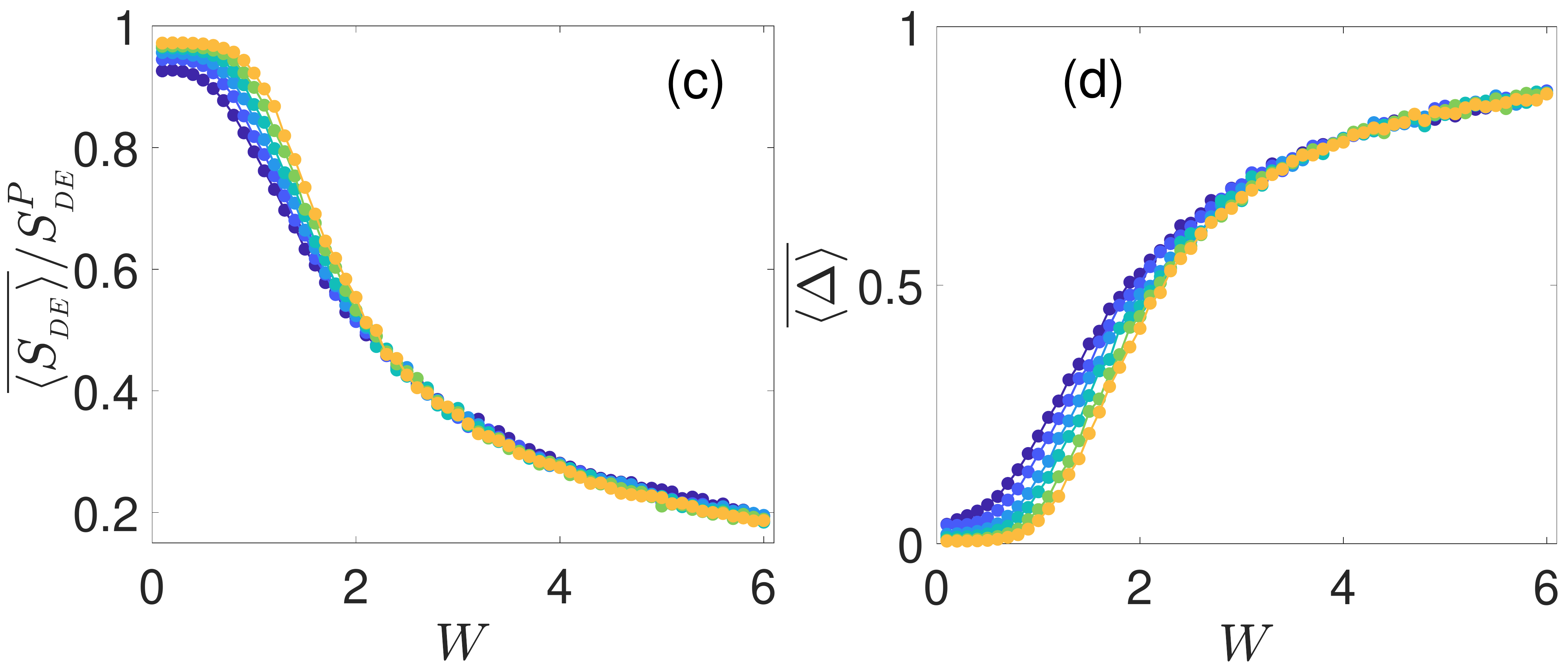}
\caption{(a): the averaged gap ratio $\overline{\langle r\rangle}$,
(b): the normalized entanglement entropy $\overline{\langle S_{_{EE}}\rangle} /S^{P}_{_{EE}}$, (c): the normalized diagonal entropy $\overline{\langle S_{_{DE}}\rangle} /S^{P}_{_{DE}}$, and (d): the Schmidt gap $\overline{\langle \Delta\rangle}$ as a function of disorder strength $W$ for different system sizes. All the quantities are obtained for long-range system with interaction's exponent $\alpha{=}0.5$ in the mid-spectrum $\varepsilon{=}0.5$.}\label{fig:Quantities_vs_W}

\end{figure}
\begin{figure*}[t!]
\includegraphics[width=\linewidth]{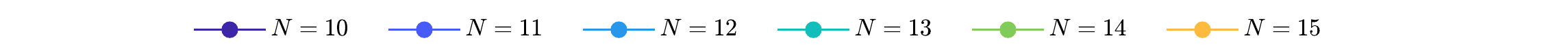}
\includegraphics[width=0.495\linewidth]{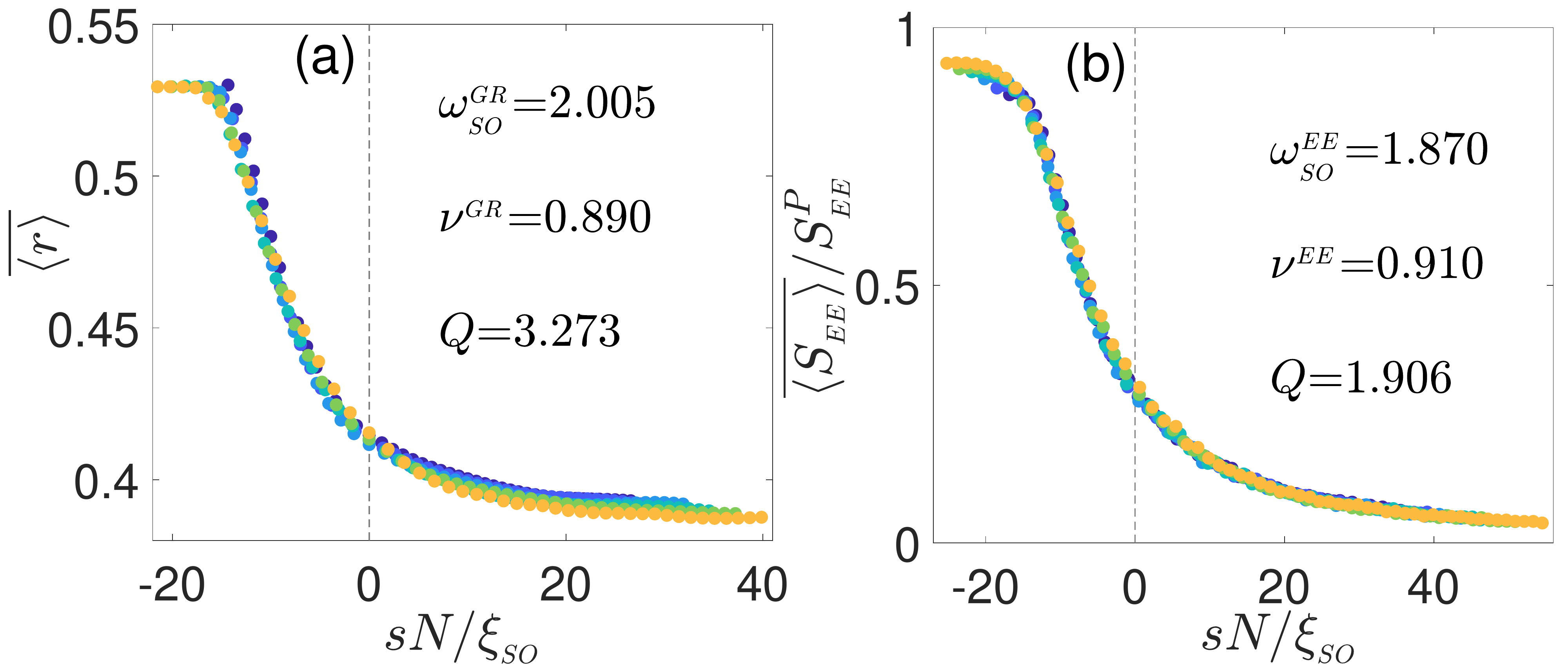}
\includegraphics[width=0.495\linewidth]{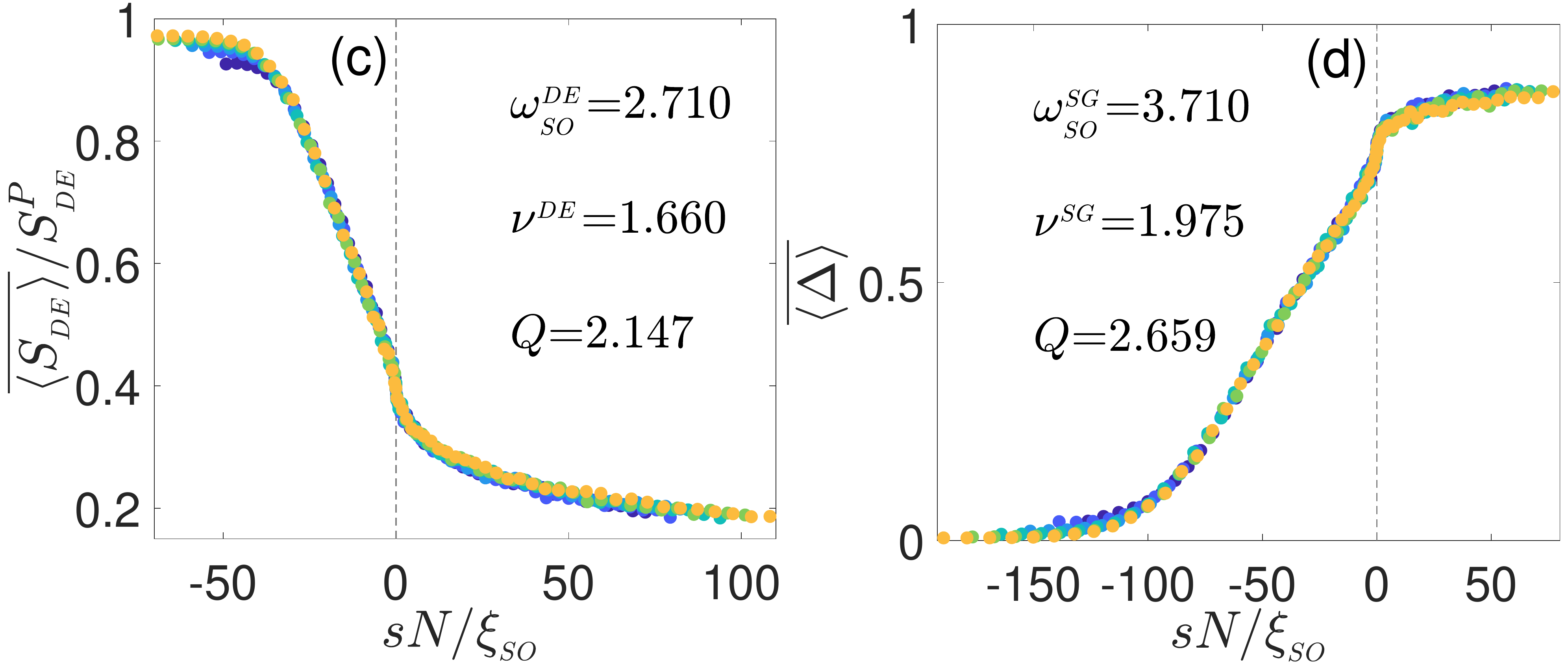}
\includegraphics[width=0.495\linewidth]{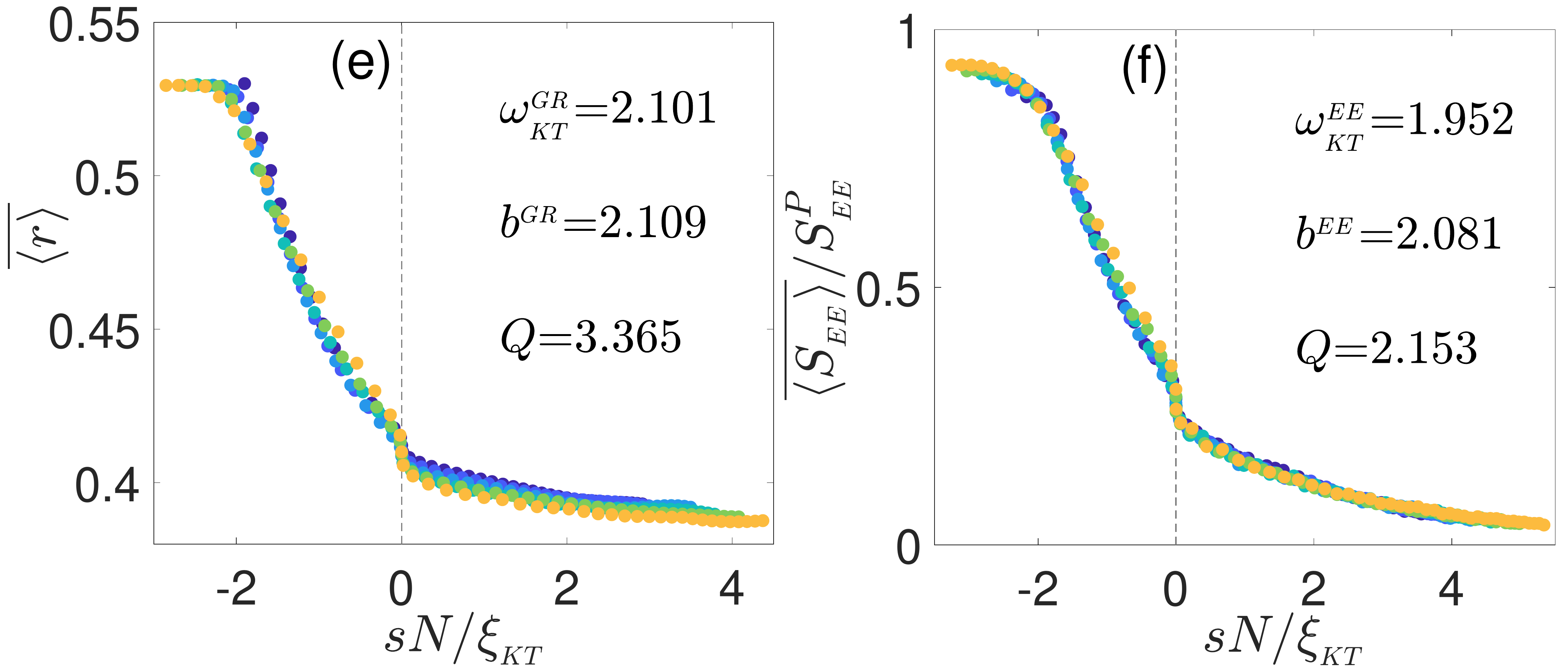}
\includegraphics[width=0.495\linewidth]{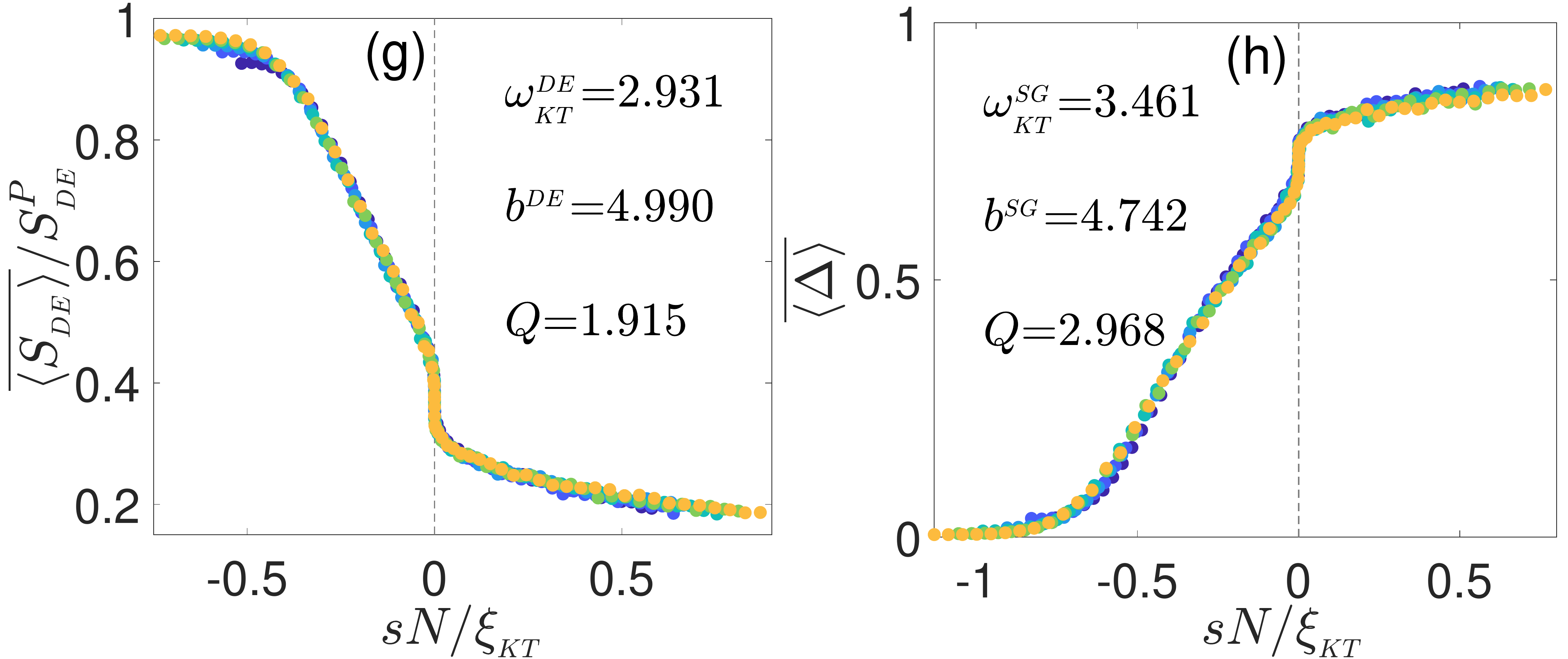}
\caption{\textbf{Upper panels}: The finite-size scaling analysis in MBL transition as a second-order phase transition. 
(a): the averaged gap ratio $\overline{\langle r\rangle}$,
(b): the normalized entanglement entropy $\overline{\langle S_{_{EE}}\rangle} /S^{P}_{_{EE}}$, 
(c): the normalized diagonal entropy $\overline{\langle S_{_{DE}}\rangle} /S^{P}_{_{DE}}$, and 
(d): the Schmidt gap $\overline{\langle \Delta\rangle}$ as a function of $sN/\xi_{_{SO}}$ with $s=sgn{(W-\omega)}$ and $\xi_{_{SO}}{\sim}\vert W- \omega\vert^{-\nu}$.
For each desired quantity $\mathcal{O}$, the  extracted critical disorder $\omega^{_{\mathcal{O}}}_{_{SO}}$, critical exponent $\nu^{_{\mathcal{O}}}$, and the quality of data collapse $Q$ have been attached on the relevant panel.
\textbf{Lower panels}: The finite-size scaling analysis in MBL transition as a Kosterlitz-Thouless transition type. 
(e): the averaged gap ratio $\overline{\langle r\rangle}$,
(f): the normalized entanglement entropy $\overline{\langle S_{_{EE}}\rangle} /S^{P}_{_{EE}}$, 
(g): the normalized diagonal entropy $\overline{\langle S_{_{DE}}\rangle} /S^{P}_{_{DE}}$, and 
(h): the Schmidt gap $\overline{\langle \Delta\rangle}$ as a function of $sN/\xi_{_{KT}}$ with $s=sgn{(W-\omega)}$ and $\xi_{_{KT}}{\sim}\exp{(b\vert W- \omega\vert^{-0.5})}$.
The extracted critical disorder $\omega^{_{\mathcal{O}}}_{_{KT}}$, fitting parameter $b^{_{\mathcal{O}}}$, and the quality of data collapse $Q$ have been attached on the relevant panel. All the data are obtained in a system with interaction's exponent $\alpha=0.5$ in the mid-spectrum $\varepsilon{=}0.5$.}\label{fig:FSS}
\end{figure*}

Regardless of the transition type, at the transition point, namely  $W{=}\omega$, the length scale diverges $\xi{\rightarrow}\infty$ (see Eqs.~(\ref{Eq.SO}), and (\ref{Eq.KT})) in the thermodynamic limit. In finite system sizes, however, this feature is revealed by $\xi{\sim}N$.  
This implies that at the  transition point the ansatz in Eq.~(\ref{eq:FSS-ansatz}) becomes independent of $N$. To observe this behavior in Figs.~\ref{fig:Quantities_vs_W} (a)-(d)
we plot the averaged gap ratio $\overline{\langle r\rangle}$, the normalized entanglement entropy $\overline{\langle S_{_{EE}}\rangle} /S^{P}_{_{EE}}$, the normalized diagonal entropy $\overline{\langle S_{_{DE}}\rangle} /S^{P}_{_{DE}}$ and the Schmidt gap $\overline{\langle \Delta\rangle}$
as a function of  disorder strength $W$ for different system sizes.
As evident in the plots, all the quantities show intersection between different lengths at around $\omega{=}2-3$.
This indeed indicates the emergence of a scale invariance behavior, of the form of Eq.~(\ref{eq:FSS-ansatz}), in the system.\\

In order to have a more precise estimation of $\omega$ and other critical exponents one has to adopt finite-size scaling analysis as a standard approach for extracting such parameters from finite-size data.
We first consider a second-order phase transition and implement the corresponding finite-size scaling analysis for all four aforementioned quantities.
In Figs.~\ref{fig:FSS}(a)-(d) we plot the averaged gap ratio $\overline{\langle r\rangle}$, the normalized entanglement entropy $\overline{\langle S_{_{EE}}\rangle} /S^{P}_{_{EE}}$, the normalized diagonal entropy $\overline{\langle S_{_{DE}}\rangle} /S^{P}_{_{DE}}$, and the Schmidt gap $\overline{\langle \Delta\rangle}$ as a function of $sN/\xi_{_{SO}}{=}sN\vert W-\omega\vert^{\nu}$ with $s{=}sgn(W-\omega)$.   
The parameters $\omega$ and $\nu$ are optimized such that the curves for different length collapse on each other. 
To achieve the best data collapse one can use an elaborate optimization scheme and minimize a proper quality function $Q$~\cite{Qfunction,Datacollaps,KT7,KT9,pyfssa1,pyfssa2},  which is defined and discussed in the Appendix.
In our case, a perfect data collapse results in $Q{=}0$ and any deviation from such perfect situation makes $Q$ larger.
In each panel of Figs.~\ref{fig:FSS}(a)-(d) we provide the optimal values of $\omega$ and $\nu$ as well as the quality function $Q$ for the resulted data collapse.
Importantly, while the averaged gap ratio $\overline{\langle r\rangle}$ and the normalized entanglement entropy $\overline{\langle S_{_{EE}}\rangle} /S^{P}_{_{EE}}$ result in $\nu{\sim}1$, consistent with previous studies~\cite{Laflorencie1,PD5,PD6,Jakub2}, the normalized diagonal entropy $\overline{\langle S_{_{DE}}\rangle} /S^{P}_{_{DE}}$ and the Schmidt gap $\overline{\langle \Delta\rangle}$ yield to exponent closer to $\nu{\sim}2$, again consistent with the previous results~\cite{Memory1,Schmidtgap,DEMBL1,DEMBL2}. 
This shows that the normalized diagonal entropy and the Schmidt gap are more consistent with Harris criterion. 
Interestingly, the value of $\omega$ which is obtained for the Schmidt gap and the diagonal entropy are remarkably higher than those for the averaged gap ratio and the entanglement entropy.
These suggest faster convergence of the Schmidt gap and the diagonal entropy towards their thermodynamic limit in comparison with the averaged gap ratio and the normalized entanglement entropy.
In other words, while the Schmidt gap and the diagonal entropy have almost reached their thermodynamic limit with system sizes of $N{=}15$, the other two quantities still show significant length dependence. \\  

Alternatively, one can also perform finite-size scaling analysis assuming that the underlying transition is of the Kosterlitz-Thouless type.      
In Figs.~\ref{fig:FSS}(e)-(h), we plot the averaged gap ratio $\overline{\langle r\rangle}$, the normalized entanglement entropy $\overline{\langle S_{_{EE}}\rangle} /S^{P}_{_{EE}}$, the normalized diagonal entropy $\overline{\langle S_{_{DE}}\rangle} /S^{P}_{_{DE}}$, and the Schmidt gap $\overline{\langle \Delta\rangle}$  as a function of $sN/\xi_{_{KT}}{=}sN\exp{(-b \vert W-\omega\vert^{-0.5})}$.
The parameters $\omega$ and $b$ are optimized such that the curves for different lengths collapse on each other. Similar to the previous case the quality of the data collapse is quantified through the same quality function $Q$.
In each panel of Figs.~\ref{fig:FSS}(e)-(h), 
we provide the optimal values of $\omega$ and $b$ as well as the quality function $Q$ for the resulted data collapse.
In the case of Kosterlitz-Thouless transition, the only relevant quantity is the transition point $\omega$.
Interestingly, similar to the previous case, the diagonal entropy and the Schmidt gap result in higher values of $\omega$ in comparison with the averaged gap ratio and the entanglement entropy. 
A remarkable observation is that the values obtained for the transition point, namely $\omega$, is very close for both types of transition.
In addition, the quality function $Q$ for all four quantities is very close for both transition types. The same issue observed in previous studies~\cite{KT7,KT9}.
The similarity of the obtained values for $\omega$ and $Q$ show 
that, indeed, one cannot discriminate between the two transitions at least based on data analysis for small system sizes.

\section{Phase diagram}
\begin{figure*}[t!]
\includegraphics[width=0.6\linewidth]{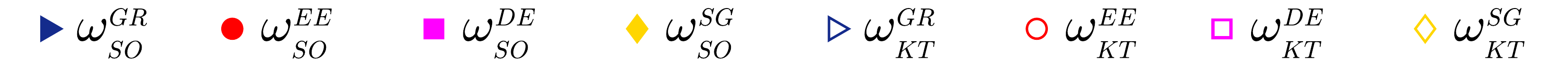}
\includegraphics[width=0.8\linewidth]{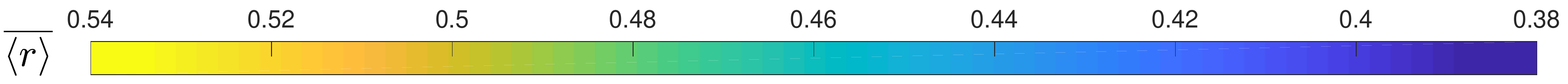}
\includegraphics[width=0.33\linewidth, height=0.25\linewidth]{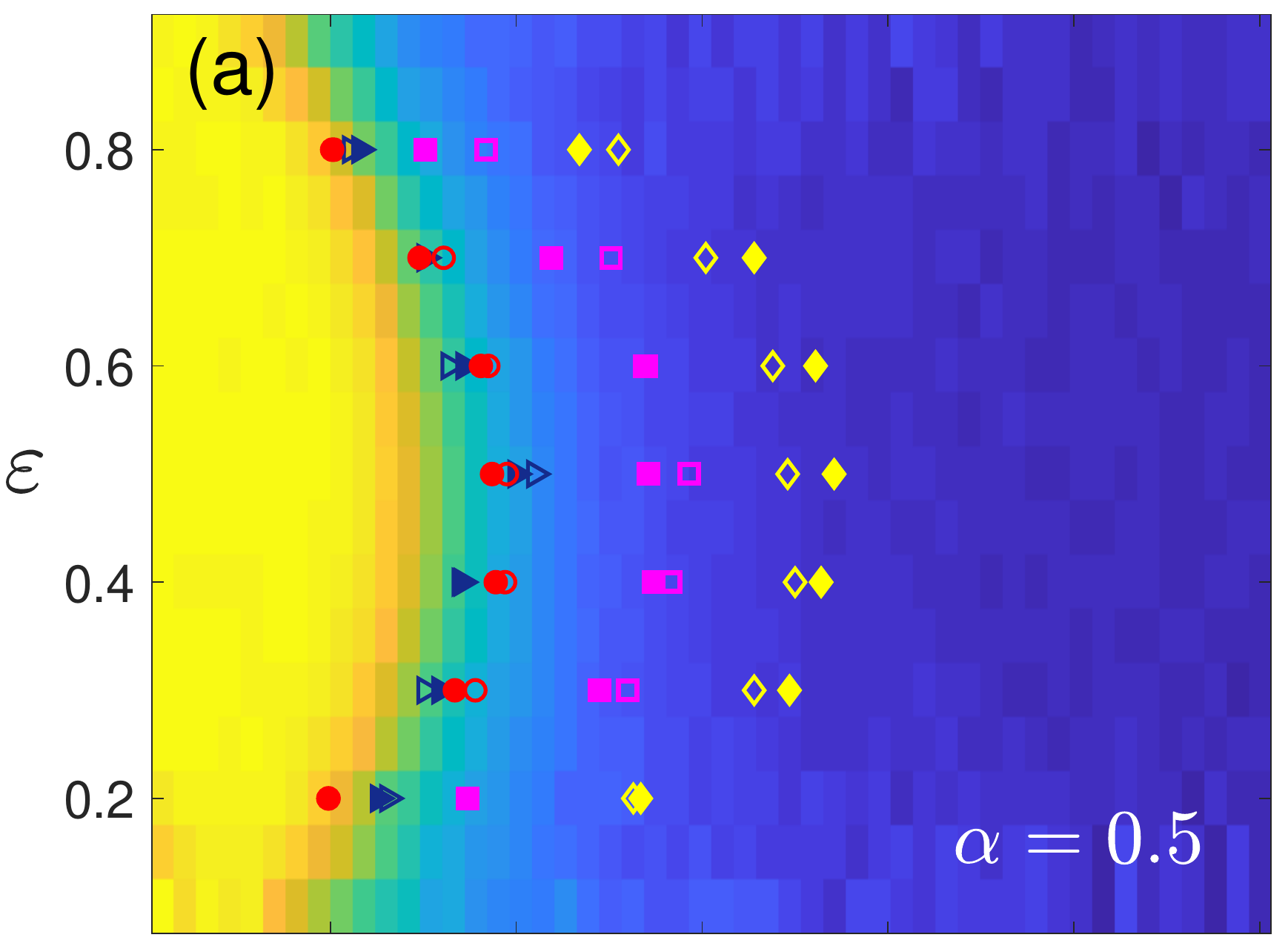}
\includegraphics[width=0.3\linewidth, height=0.25\linewidth]{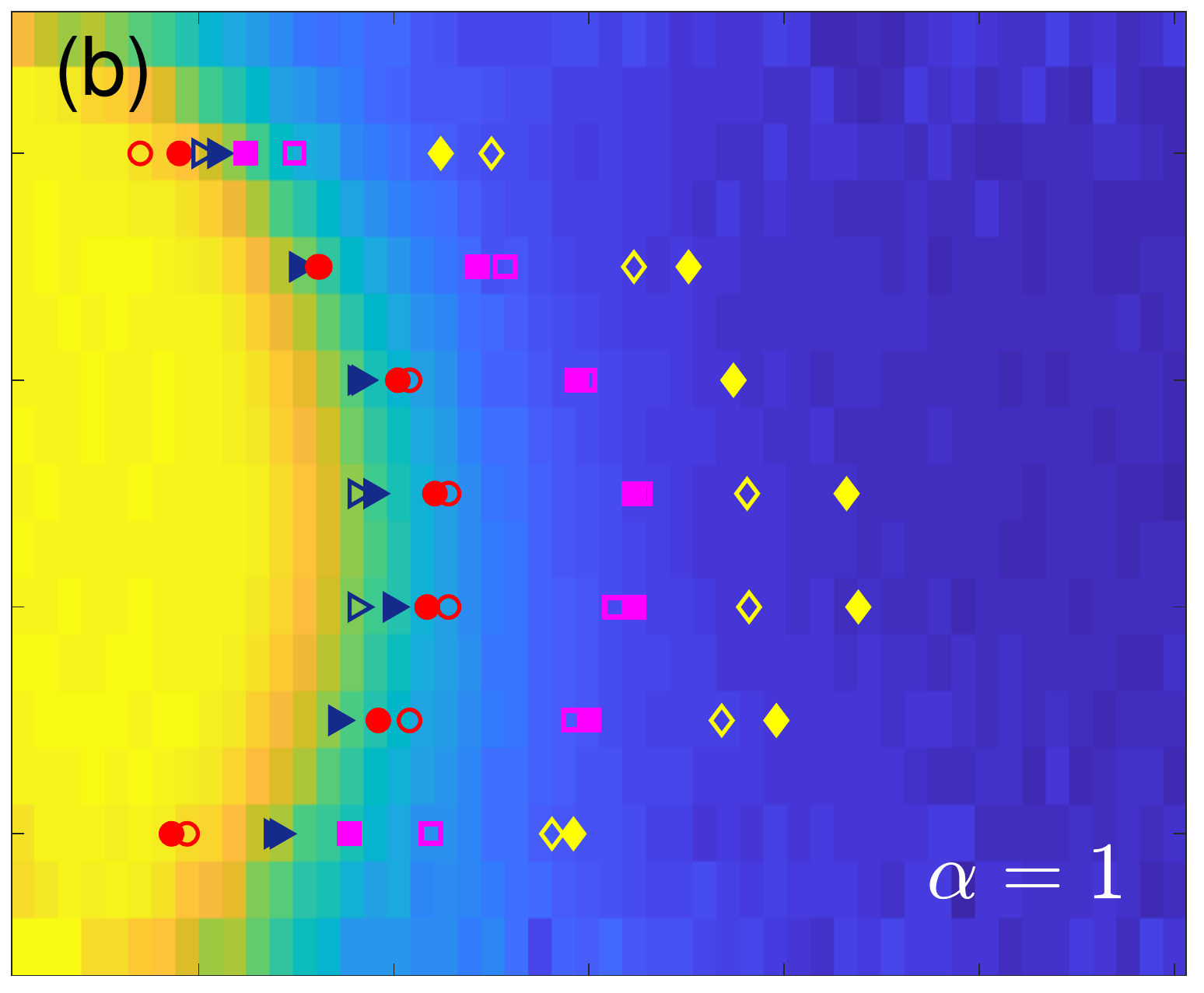}
\includegraphics[width=0.33\linewidth, height=0.25\linewidth]{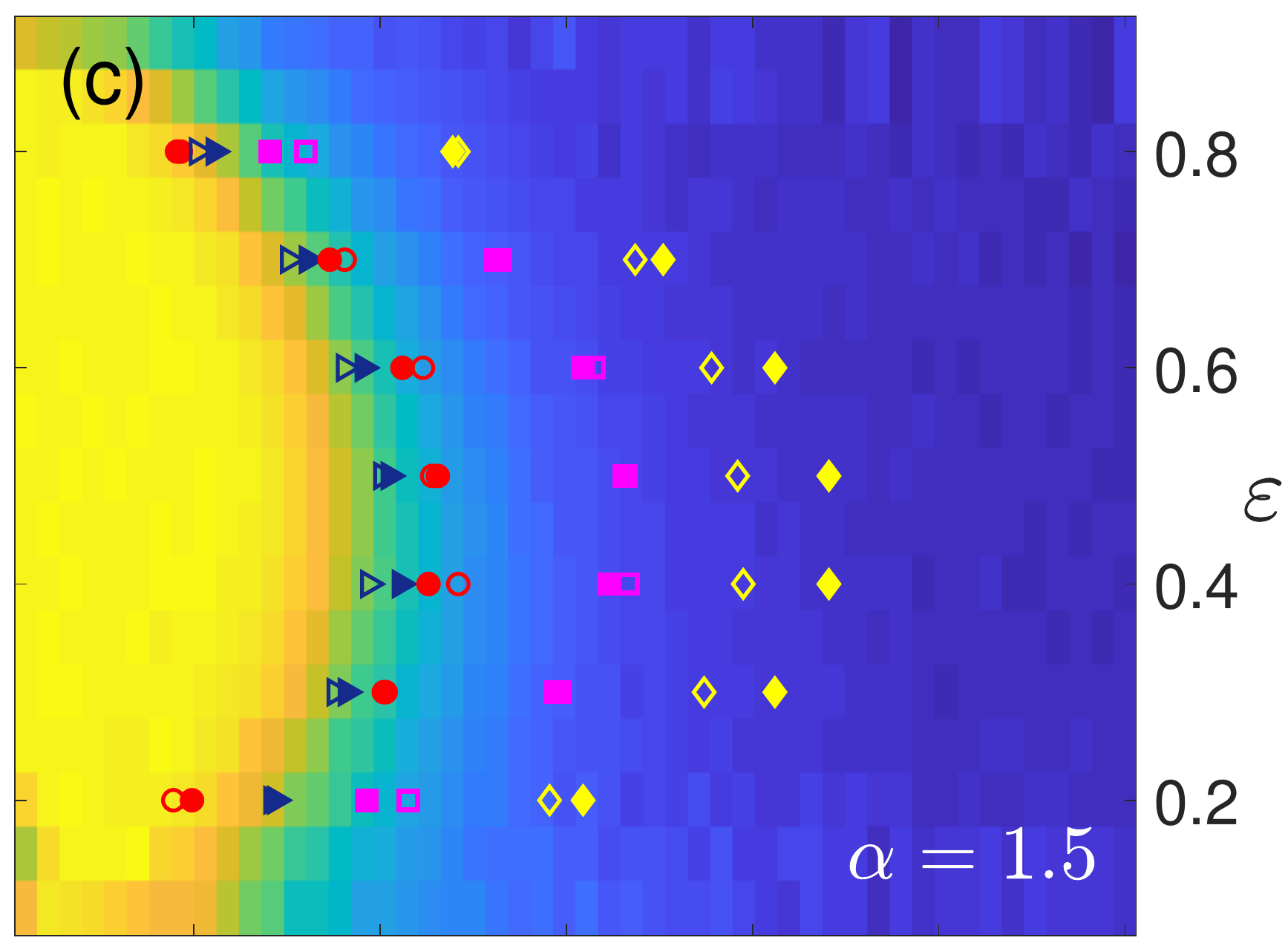}
\includegraphics[width=0.33\linewidth, height=0.30\linewidth]{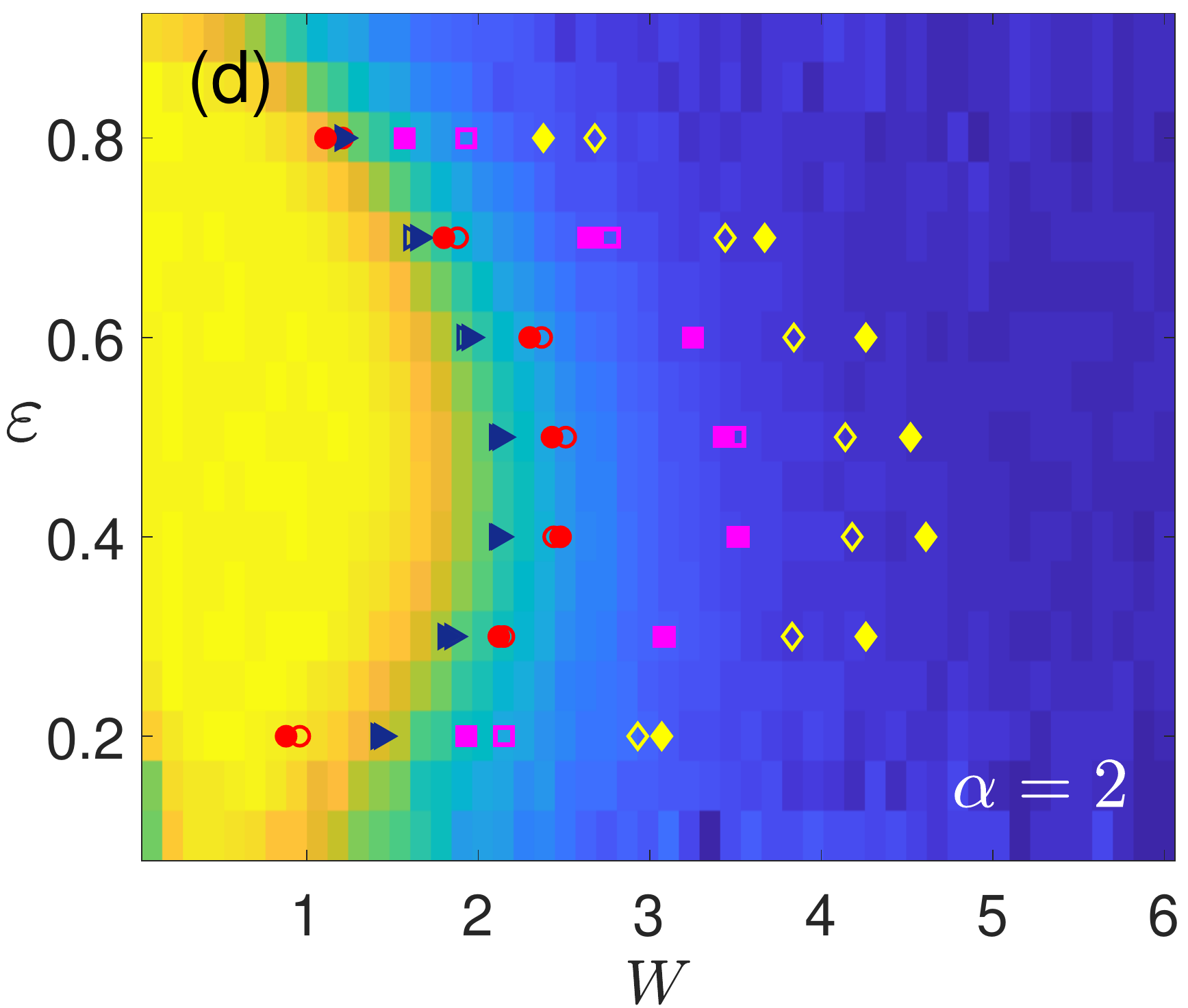}
\includegraphics[width=0.3\linewidth, height=0.30\linewidth]{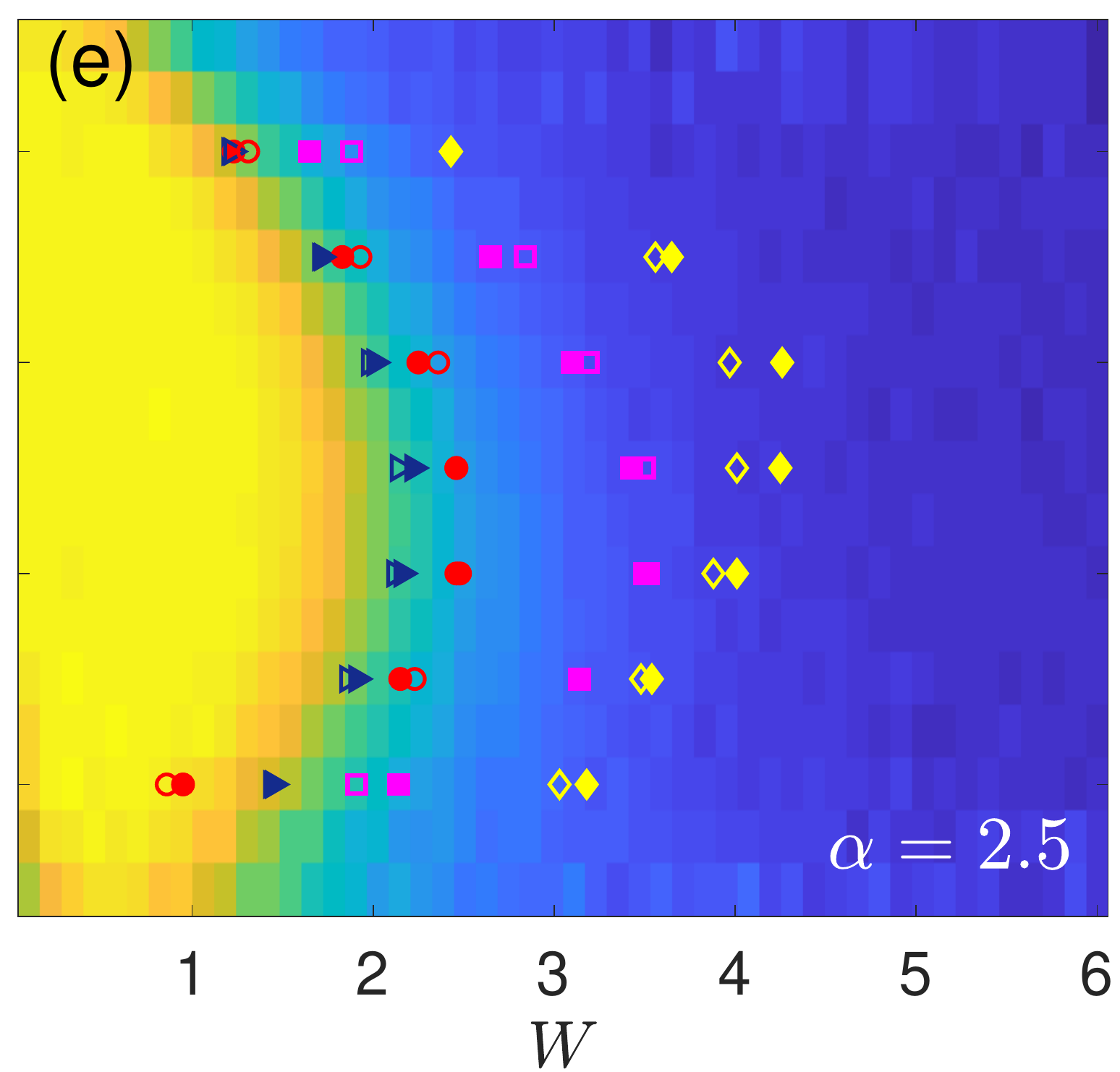}
\includegraphics[width=0.33\linewidth, height=0.30\linewidth]{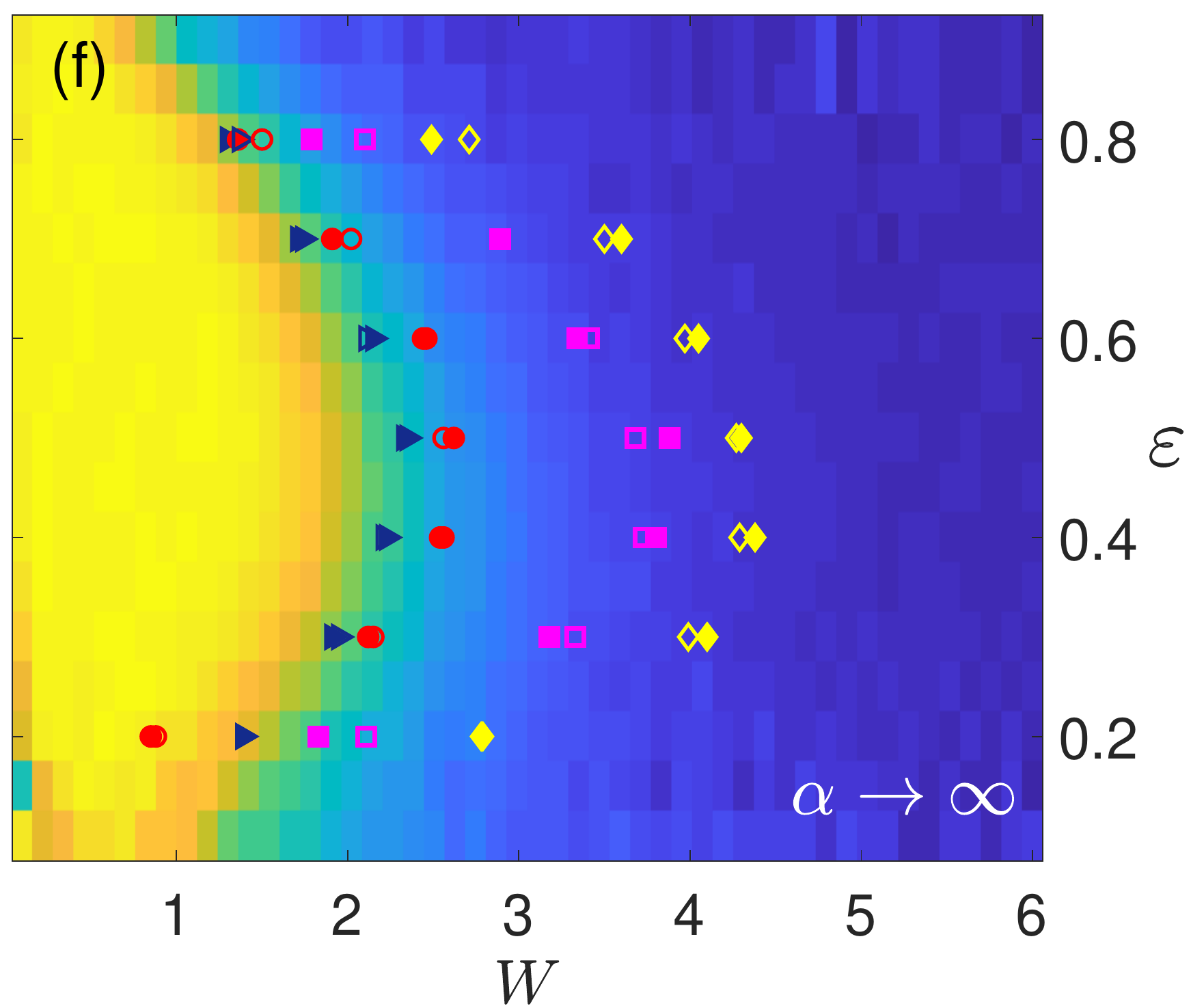}
\caption{The averaged gap ratio $\overline{\langle r\rangle}$ as function of energy $\varepsilon$ and disorder strength $W$ for different interaction ranges from (a): $\alpha{=}0.5$ to (f): $\alpha{\rightarrow}\infty$.
The extracted critical disorders $\omega^{_{\mathcal{O}}}_{_{SO}}$ and $\omega^{_{\mathcal{O}}}_{_{KT}}$ are obtained by considering the transition as second-order and Kosterlitz-Thouless type, respectively. In all the panels, $\omega^{_{GR}}_{_{SO,KT}}$, $\omega^{_{EE}}_{_{SO,KT}}$, $\omega^{_{DE}}_{_{SO,KT}}$, and $\omega^{_{SG}}_{_{SO,KT}}$ are obtained from finite-size scaling analysis for the averaged gap ratio $\overline{\langle r\rangle}$, the normalized entanglement entropy $\overline{\langle S_{_{EE}}\rangle} /S^{P}_{_{EE}}$, the normalized diagonal entropy $\overline{\langle S_{_{DE}}\rangle} /S^{P}_{_{DE}}$, and the Schmidt gap $\overline{\langle\Delta\rangle}$, respectively.  }\label{fig:PD}
\end{figure*}

In this section, we extend the finite-size scaling analysis for all energy spectrum $\varepsilon$ and various values of $\alpha$.
To achieve this, at a desired energy $\varepsilon$ we pick $M=50$ eigenstates as depicted in Figs.~\ref{fig:Dos} (c)-(d). 
For any given $\alpha$ and energy scale $\varepsilon$, 
we perform finite-size scaling analysis by considering both second-order and  Kosterlitz-Thouless transition types and extract the corresponding transition points and critical exponents.    
In Figs.~\ref{fig:PD}(a)-(f) we plot  the averaged gap ratio $\overline{\langle r \rangle}$ as a function of energy $\varepsilon$  and  disorder strength $W$ for $N{=}15$ and various values of $\alpha$.
By increasing the disorder strength, one can continuously move from the ergodic phase (light region) to the localized phase (dark region).
By considering a quantity $\mathcal{O}$, which is chosen to be any of the four quantities introduced before, at different energy scales we can use either second-order or Kosterlitz-Thouless ansatzes
 to determine the transition points $\omega^{_{\mathcal{O}}}_{_{SO}}$ and $\omega^{_{\mathcal{O}}}_{_{KT}}$, respectively. 
By focusing on the second-order transition, 
in Figs.~\ref{fig:PD}(a)-(f) we present the transition points  $\omega^{_{GR}}_{_{SO}}$ (filled solid triangles), $\omega^{_{EE}}_{_{SO}}$ (filled solid circles), $\omega^{_{DE}}_{_{SO}}$ (filled solid squares), and $\omega^{_{SG}}_{_{SO}}$ (filled solid diamonds) obtained from finite-size scaling analysis for the averaged gap ratio $\overline{\langle r\rangle}$, the normalized entanglement entropy $\overline{\langle S_{_{EE}}\rangle} /S^{P}_{_{EE}}$, the normalized diagonal entropy $\overline{\langle S_{_{DE}}\rangle} /S^{P}_{_{DE}}$, and the Schmidt gap $\overline{\langle\Delta\rangle}$, respectively.
Similarly, one can perform finite-size scaling analysis for the four quantities using Kosterlitz-Thouless transition ansatz.
The results are also shown in Figs.~\ref{fig:PD}(a)-(f) in which the transition points   $\omega^{_{GR}}_{_{KT}}$ (empty triangles), $\omega^{_{EE}}_{_{KT}}$ (empty circles), $\omega^{_{DE}}_{_{KT}}$ (empty squares), and $\omega^{_{SG}}_{_{KT}}$ (empty diamonds) are depicted across the whole spectrum. \\

Four main features can be observed from Fig~\ref{fig:PD}. 
First, all considered quantities, regardless of the transition type, reveal a generic D-shape phase boundary along the spectrum, the well-known mobility edge.
While this feature has already been observed for MBL transition as a second-order phase transition, both analytically~\cite{Laflorencie1,Area-law1,
Mobility1,Mobility2,Mobility3,Mobility4,
Mobility5,Mobility6,Mobility7,Mobility8,Mobility9}
 and experimentally~\cite{superconducting1}, 
the emergence of this feature for the Kosterlitz-Thouless transition has not been reported previously. 
Second, decreasing $\alpha$ (i.e. making the interaction more long-range) enhances the localization power such that the localization occurs for smaller values of disorder.
This can be describe by the fact that every spin configuration of the chain induces an effective Zeeman energy splitting at a given site.
Thus, the superposition of different spin configurations can play like an effective random field and thus enhances the localization power which is in agreement with previous works~\cite{Yousefjani}.
Third, apart from Schmidt gap, the other three quantities result in very close transition points for both second-order and Kosterlitz-Thouless transition types.
This shows that the detection of the MBL phase is very robust and hardly depends on the type of the transition. 
Forth, one can clearly observe a hierarchy among the four quantities for revealing the MBL transition point
\begin{equation}\label{Eq:W}
   \omega^{_{SG}}_{_{SO,KT}} \geqslant \omega^{_{DE}}_{_{SO,KT}} 
   \geqslant \omega^{_{EE}}_{_{SO,KT}} \geqslant \omega^{_{GR}}_{_{SO,KT}}.
\end{equation}
The reason that each quantity gives a distinct transition point is  indeed a finite-size effect indicating different convergence speed towards their thermodynamic limit.\\

\begin{figure}[t]
\includegraphics[width=0.5\linewidth]{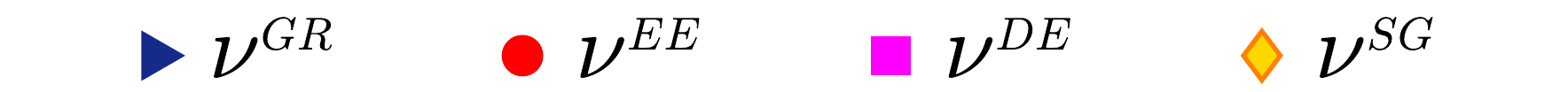}
\includegraphics[width=\linewidth]{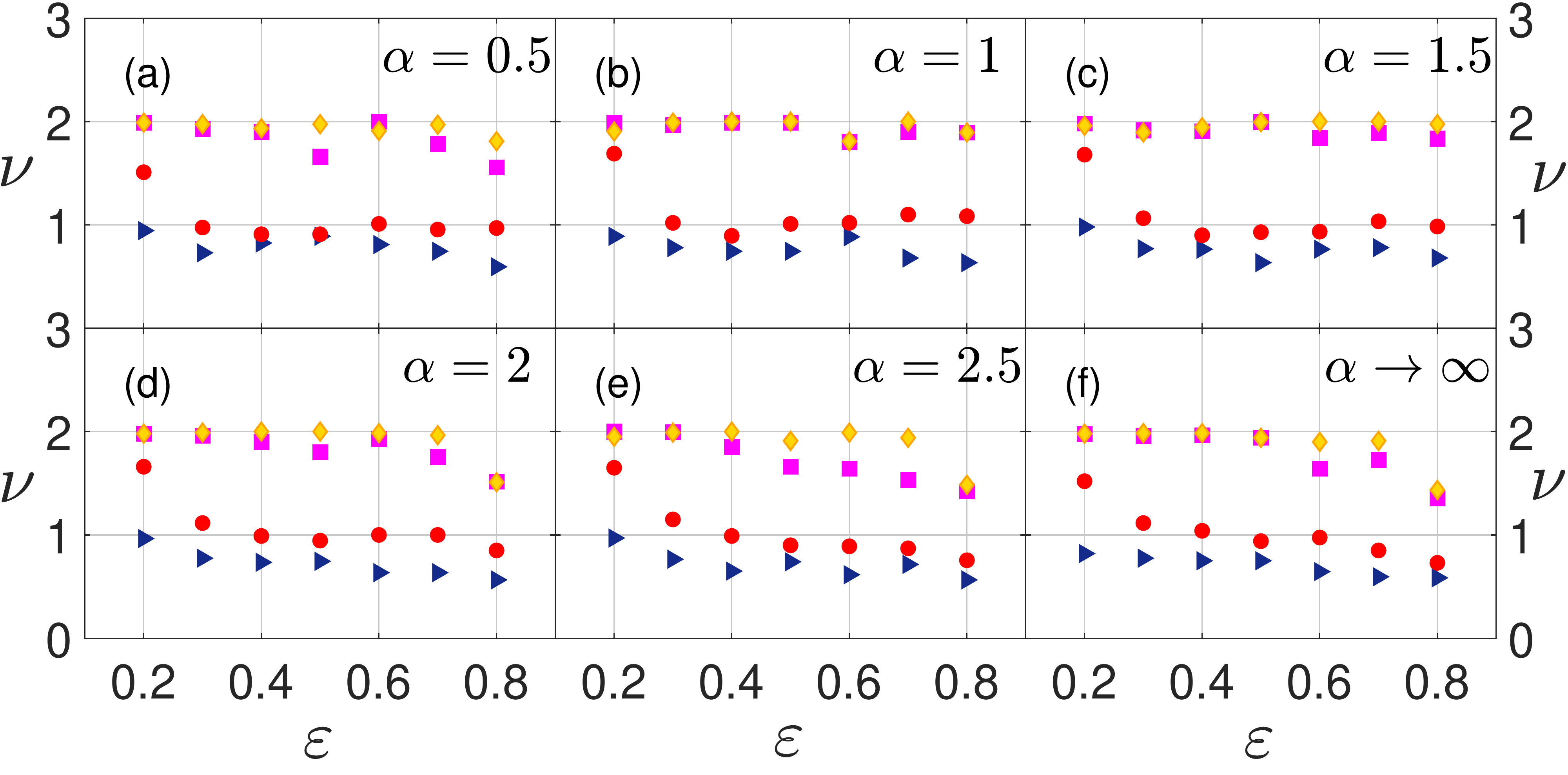}
\caption{The critical exponent $\nu$'s (see Eq.~(\ref{Eq.SO})) as a function of energy scale $\varepsilon$ for various interaction ranges from (a): $\alpha{=}0.5$ to (f): $\alpha{\rightarrow}\infty$. 
These exponents are obtained from finite-size scaling analysis for different observables including the averaged gap ratio $\overline{\langle r\rangle}$ (dark blue triangle), the normalized entanglement entropy $\overline{\langle S_{_{EE}}\rangle} /S^{P}_{_{EE}}$ (red circle), the normalized diagonal entropy $\overline{\langle S_{_{DE}}\rangle} /S^{P}_{_{DE}}$ (magenta square), and the Schmidt gap $\overline{\langle \Delta\rangle}$ (yellow diamond). }\label{fig:nu}
\end{figure}

As mentioned before, finite-size scaling analysis not only provides the transition point $\omega$, but also reveals other exponents such as $\nu$ (for the second-order transition) and $b$ (for the Kosterlitz-Thouless transition).   
In Figs.~\ref{fig:nu}(a)-(f) we plot the critical exponent $\nu$ obtained from finite-size scaling analysis of all four quantities as a function of rescaled energy $\varepsilon$ for different interaction range $\alpha$.
For all values of $\alpha$ and $\varepsilon$ one can observe  a hierarchy among the four quantities, namely  
\begin{equation}\label{Eq:nu}
\nu^{_{SG}} \geqslant  \nu^{_{DE}} \geqslant \nu^{_{EE}} \geqslant \nu^{_{GR}}.
\end{equation} 
By comparing Eqs.~(\ref{Eq:W}) and (\ref{Eq:nu}), one can see that the larger value of the transition point $\omega$ implies a larger value for the critical exponent $\nu$. 
Moreover, the critical exponent $\nu$ obtained from the averaged gap ratio $\nu^{_{GR}}$ and the entanglement entropy $\nu^{_{EE}}$ 
always remain near one, i.e. $\nu{\sim}1$, showing the usual contradiction with the Harris bound. 
On the other hand, the critical exponent obtained from the diagonal entropy ($1.5{<}\nu^{_{DE}}{\leqslant}2$) and the Schmidt gap ($\nu^{_{SG}}{\cong}2$) are far more consistent with the Harris bound.  
This shows that if the underlying transition is second-order then diagonal entropy and Schmidt gap indeed have better convergence towards their thermodynamic limits even for system sizes as small as $N{=}15$.
Schmidt gap only depends on the two largest eigenvalues of the reduced density matrix of the half system and ignores the rest of the spectrum.  
Similarly, in diagonal entropy 
one also relies on partial information through  setting the off-diagonal terms of the half system reduced density matrix to zero (in computational basis). 
This artificial decohering action has recently been proposed for emulating the thermodynamic behavior in MBL context~\cite{Memory1}. 
In fact, in the deep ergodic and MBL phases,  the reduced density matrix is expected to be decohered in the thermodynamic limit due to being either maximally mixed (i.e. maximally entangled with the rest of the system) or a product pure state (i.e. fully separable), respectively. 
Note that, the variation of $\nu$ at different rescaled energy $\varepsilon$ is a consequence of several anti-crossings between different energy eigenstates. \\

\begin{figure}[t!]
\includegraphics[width=0.45\linewidth]{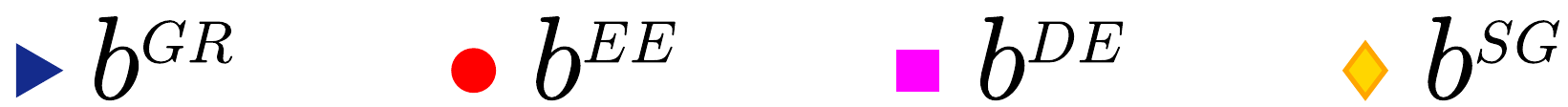}
\includegraphics[width=\linewidth]{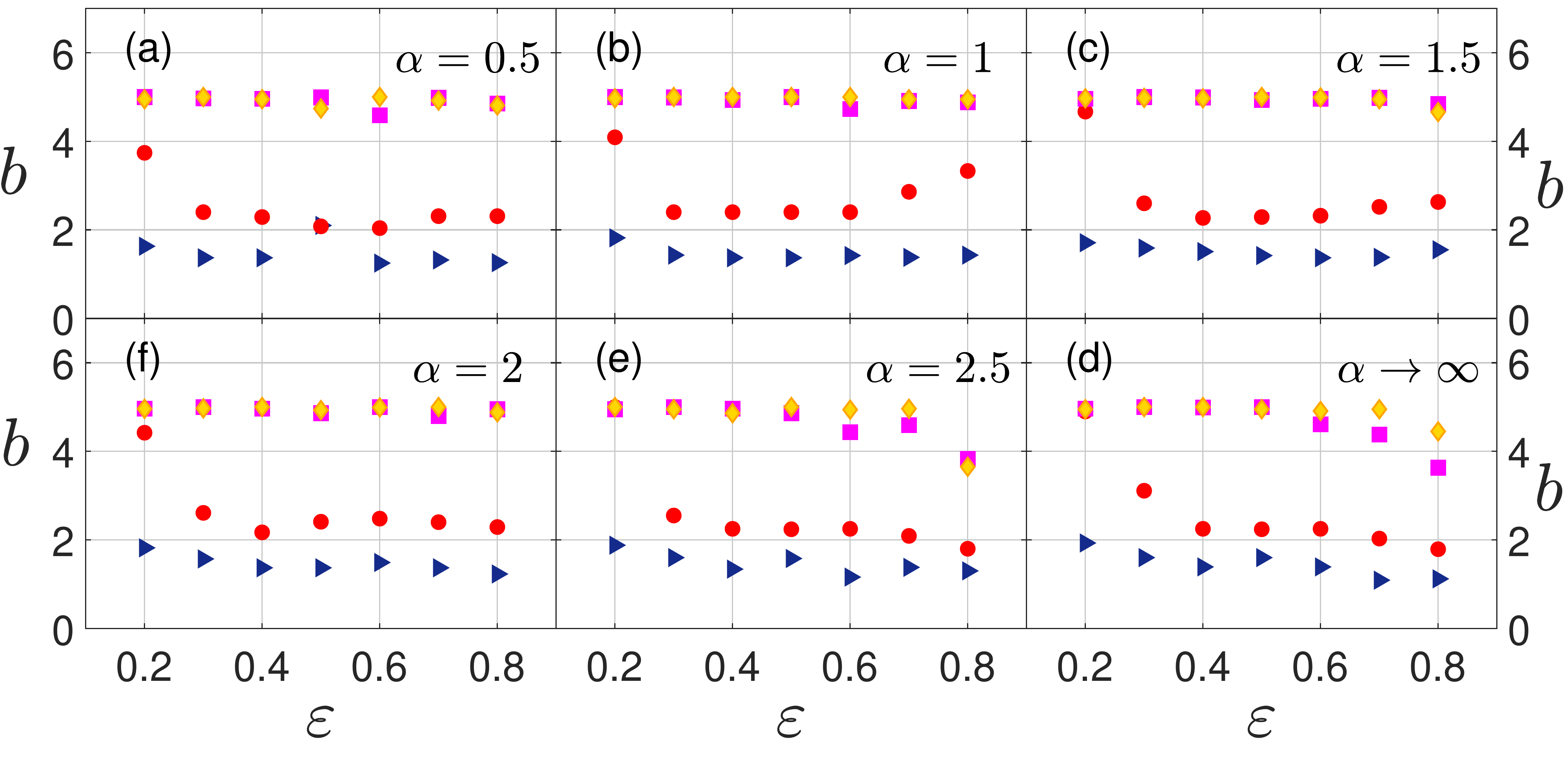}
\caption{The fitting parameter $b$ for Kosterlitz-Thouless transition type (see Eq.~(\ref{Eq.KT})) as a function of energy scale $\varepsilon$ for various interaction ranges from (a): $\alpha{=}0.5$ to (f): $\alpha{\rightarrow}\infty$. 
These parameters are obtained from finite-size scaling analysis for different observables including the averaged gap ratio $\overline{\langle r\rangle}$ (dark blue triangle), the normalized entanglement entropy $\overline{\langle S_{_{EE}}\rangle} /S^{P}_{_{EE}}$ (red circle), the normalized diagonal entropy $\overline{\langle S_{_{DE}}\rangle} /S^{P}_{_{DE}}$ (magenta square), and the Schmidt gap $\overline{\langle \Delta\rangle}$ (yellow diamond). }\label{fig:b}
\end{figure}

A side product of finite-size scaling analysis, using Kosterlitz-Thouless ansatz, is the non-universal fitting parameter $b$, see Eq.~(\ref{Eq.KT}). 
In Figs.\ref{fig:b} (a)-(f), we present the optimal values of $b$  obtained from all the four quantities  as a function of rescaled energy $\varepsilon$  for various values of $\alpha$. 
As it is evident from the figure, the quantities preserve the same hierarchy for the values of $b$, namely
\begin{equation}\label{Eq:b}
b^{_{SG}} \geqslant  b^{_{DE}} \geqslant b^{_{EE}} \geqslant b^{_{GR}}.
\end{equation} 
\\
 
Finely, we have to emphasize that our numerical analysis does not favor  
any of the transition types for describing the MBL transition. 
This is indicated in the quality function $Q$.
While at a certain energy the second-order transition may show slightly better $Q$, in another energy it is the opposite. 
Therefore, due to the small accessible system sizes, our study cannot conclusively settle the type of the MBL transition.

\section{Conclusion} 
In this paper, we have determined the phase boundary between ergodic and MBL phases along the energy spectrum, i.e. the mobility edge, for a Heisenberg spin-chain with long-range interaction in the presence of disordered magnetic field. 
This has been done using several quantities, including the level statistics ratio, the entanglement entropy, the diagonal entropy, and the Schmidt gap.
We showed that long-range interaction enhances the localization effect and shifts the mobility edge in favor of MBL.
Two types of transition, namely second-order and Kosterlitz-Thouless, have been investigated for describing the emergence of the MBL phase.
Our analysis show that the transition type does not change the description of the mobility edge and the phase boundary is hardly affected by that.
However, the choice of the quantity may indeed shift the mobility edge significantly. 
This is a finite-size effect 
which is related to different convergence speed of the quantities in reaching their thermodynamic limit.
We have established a hierarchy among the quantities with respect to their corresponding transition points and critical exponents. 
In the whole spectrum, larger values of the transition point $\omega$ implies larger values for the critical exponents, $\nu$ and $b$. 
Schmidt gap and diagonal entropy seem to converge faster towards their thermodynamic limit such that in the case of second-order transition, their critical exponent $\nu$ show   significant consistency with the Harris bound, for a wide range of the long-range interactions.  
The closeness of the quality functions obtained for both second-order and Kosterlitz-Thouless transitions, shows that one cannot conclusively determine the transition type using small system sizes considered in our analysis.

\section{Acknowledgment}

A.B. acknowledge support from the National Key R\&D Program of China (Grant No. 2018YFA0306703), the National Science Foundation of China (Grants No. 12050410253, No. 92065115, and No. 12274059), and the Ministry of Science and Technology of China (Grant No. QNJ2021167001L). R.Y. thanks the National Science Foundation of China for the International Young Scientists Fund (Grant No. 12250410242). 

\appendix* 
\setcounter{equation}{0}
\setcounter{figure}{0}
\setcounter{table}{0}
\renewcommand{\theequation}{A\arabic{equation}}
\renewcommand{\thefigure}{A\arabic{figure}}
\renewcommand{\thetable}{A\arabic{table}}
\section{}

In the main text, we discussed that for both second-order and Kosterlitz-Thouless transition types, a more precise estimation of the transition point $\omega$  and other relevant exponents, namely $\nu$ and $b$, can be obtained by adopting a standard finite-size scaling analysis.  
In this method, by rescaling data as a function of $sN/\xi_{_{OS}}=sN\vert W-\omega\vert^{\nu}$ for the second-order transition or $sN/\xi_{_{KT}}=sN\exp{(-b\vert W-\omega\vert^{-0.5})}$ for the Kosterlitz-Thouless transition type, one can collapse all the curves on each other. 
For achieving the best data collapse one can use the optimization algorithms for minimizing a quality function $Q$. 
In this work, following the Refs.~\cite{KT7,KT9}, we select 
\begin{equation}\label{Eq.Q}
Q = \dfrac{ \sum_{m=1}^{\mathcal{N}-1} \vert X_{m+1} - X_{m}\vert}{\max{(X)}-\min{(X)}} -1,
\end{equation}
as our quality function, which according to our experience seemed to be stabler and more precise. 
Here, $X$ stands for the desired quantity with $\mathcal{N}$ values at different disorder strength $W$ and system size $N$ which has been sorted according to non-decreasing values of $sgn(\vert W-\omega\vert)N/\xi$. 
This quality function converges to zero in the case of an ideal data collapse to a single curve and  larger otherwise.


\end{document}